\documentclass[12pt]{article}

\author{Yu.~M.~Zinoviev
       \thanks{E-mail address: yurii.zinoviev@ihep.ru} \\
        {\it Institute for High Energy Physics} \\
        {\it Protvino, Moscow Region, 142280, Russia}}
\title{Massive $N=1$ supermultiplets \\ with arbitrary superspins}

\date{}

\begin{document}

\maketitle

\begin{abstract}
In this paper we give explicit construction of massive $N=1$
supermultiplets in flat $d=4$ Minkowski space-time. We work in a
component on-shell formalism based on gauge invariant description of
massive integer and half-integer spin particles where massive
supermultiplets are constructed out of appropriate set of massless
ones.
\end{abstract}

\thispagestyle{empty}
\newpage
\setcounter{page}{1}

\section*{Introduction}

In a flat space-time massive spin $s$ particles in a massless limit
decompose into massless spin $s$, $s-1$, $\dots$ ones. This,
in particular, leads to the possibility of gauge invariant
description of massive spin $s$ particles, e.g.
\cite{Zin83}-\cite{BKR07}. In this, two different approaches could be
used. From one hand, one can start with usual non gauge invariant
description of massive particle and achieve gauge invariance through
the introduction of additional fields (thus promoting second class
constraints into the first class ones). From the other hand, one can
start with the appropriate set of massless particles having gauge
invariance from the very beginning and obtain massive particle 
description as a deformation of massless theory. This last approach
closely mimic situation in spontaneous gauge symmetry breaking where
gauge field has to eat some Goldstone field(s) to become massive.

In the supersymmetric theories all particles must belong to some
supermultiplet, massive or massless. Till now most of investigations
in supersymmetric theories where bounded to massless supermultiplets.
Quite a few results on massive supermultiplets mainly devoted to
superspins 1 and 3/2 exist \cite{BGPL02}-\cite{Zin07}. The aim of this
paper is to extend these results to include massive $N=1$
supermultiplets with arbitrary superspins. Certainly, it would be nice
to have superfield off-shell description of such supermultiplets, but
as previous results clearly show it is a highly non-trivial task. So
in this paper we restrict ourselves with component on-shell formalism
in terms of physical fields. The same reasoning on the massless limit
means that massive supermultiplets could (should) be constructed out
of the massless ones in the same way as massive particles out of the
massless ones. So our approach will be supersymmetric generalization
of the second approach to massive particle description mentioned
above. Namely, we will start with appropriate set of massless
supermultiplets and obtain massive one as a smooth deformation.

The paper is organized as follows. Though our previous examples on
massive superspin 1 \cite{Zin07} and superspin 3/2 \cite{Zin02}
supermultiplets already give important hints on how general case of
arbitrary superspin could looks like, due to peculiarities of
lower spin fields they are not enough to achieve such generalization.
Thus, in the first two sections we give two more concrete examples,
namely massive supermultiplets with superspin 2 and 5/2,
correspondingly. All these and subsequent results heavily depend on
the gauge invariant description of massive particles with integer
\cite{KZ97,Zin01} and half-integer \cite{Met06} spins as well as on
the known form of massless supermultiplets \cite{Cur79}. For reader
convenience and to make paper self-contained, in the next two sections
we give all necessary formulas in compact condensed notations. One of
the lessons from previous investigations is that the structures of
massive supermultiplets with integer and half-integer superspins are
different, so in the last two sections we consider these two cases
separately. We will see that, in spite of large number of fields, all
calculations are pretty straightforward and mainly combinatorical.

\section{Superspin 2}

Massive superspin 2 supermultiplet contains four massive particles
with spins 5/2, 2, 2' and 3/2, correspondingly. In the massless limit
massive supermultilets must decompose into the appropriate set of
massless ones in the same way as massive spin s particles --- into
massless spin s, s-1, ... ones. Simple counting of physical degrees
of freedom immediately gives:
$$
\left( \begin{array}{ccc}  & 5/2 & \\ 2 &  & 2' \\  & 3/2 &
\end{array} \right) \quad \Rightarrow \quad
\left( \begin{array}{c} 5/2 \\ 2 \end{array} \right) \oplus
\left( \begin{array}{c} 2 \\ 3/2 \end{array} \right) \oplus
\left( \begin{array}{c} 3/2 \\ 1 \end{array} \right) \oplus
\left( \begin{array}{c} 1 \\ 1/2 \end{array} \right) \oplus
\left( \begin{array}{c} 1/2 \\ 0,0' \end{array} \right)
$$
So we will start with five massless supermultiplets $(\Phi_{\mu\nu},
h_{\mu\nu})$, $(f_{\mu\nu}, \Psi_\mu)$, $(\Omega_\mu, A_\mu)$,
$(B_\mu, \psi)$ and $(\chi, z)$. From our previous experience with
massive superspin 1 and superspin 3/2 supermultiplets we know that it
is crucial for the construction of massive supermultiplets to make
dual rotation of vector $A_\mu$ and axial-vector $B_\mu$ fields
mixing massless supermultiplets containing these fields. But now we
have two tensor fields $h_{\mu\nu}$ and $f_{\mu\nu}$ as well,
moreover they necessarily must be tensor and pseudo-tensor ones. Thus
we have to consider the possibility to mix massless supermultiplets
with these fields as well and the real structure of massless
supermultiplets we are going to work with looks like:
$$
\left( \begin{array}{ccc}  & \Phi_{\mu\nu} & \\ h_{\mu\nu} & &
f_{\mu\nu} \\  & \Psi_\mu & \end{array} \right) \quad \oplus \quad
\left( \begin{array}{ccc}  & \Omega_\mu & \\ A_\mu &  & B_\mu \\  &
\psi & \end{array} \right) \quad \oplus \quad
\left( \begin{array}{c} \chi \\ z \end{array} \right)
$$
Then, introducing a sum of the massless Lagrangians for bosonic
fields:
\begin{eqnarray}
{\cal L}_0 &=& \frac{1}{2} \partial^\alpha h^{\mu\nu} \partial_\alpha
h_{\mu\nu} - (\partial h)^\mu (\partial h)_\mu + (\partial h)^\mu
\partial_\mu h - \frac{1}{2} \partial^\mu h \partial_\mu h + \nonumber
\\
 && + \frac{1}{2} \partial^\alpha f^{\mu\nu} \partial_\alpha
f_{\mu\nu} - (\partial f)^\mu (\partial f)_\mu + (\partial f)^\mu
\partial_\mu f - \frac{1}{2} \partial^\mu f \partial_\mu f - \nonumber
\\
 && - \frac{1}{4} A_{\mu\nu}{}^2 - \frac{1}{4} B_{\mu\nu}{}^2 +
\frac{1}{2} (\partial_\mu \varphi)^2 + \frac{1}{2} (\partial_\mu
\pi)^2
\end{eqnarray}
as well as sum of the massless Lagrangians for fermionic fields:
\begin{eqnarray}
{\cal L}_0 &=& \frac{i}{2} \bar{\Phi}^{\mu\nu} \hat{\partial}
\Phi_{\mu\nu} - 2 i (\bar{\Phi}\gamma)^\mu (\partial\Phi)_\mu 
+ i (\bar{\Phi}\gamma)^\mu \hat{\partial} (\gamma\Phi)_\mu +
i (\bar{\Phi}\gamma\partial) \Phi - \frac{i}{4}
\bar{\Phi} \hat{\partial} \Phi - \nonumber  \\
 && - \frac{i}{2} \bar{\Psi}^\mu \hat{\partial} \Psi_\mu + i
(\bar{\Psi}\gamma) (\partial\Psi) - \frac{i}{2}
(\bar{\Psi}\gamma) \hat{\partial} (\gamma\Psi) + \frac{i}{2}
\bar{\psi} \hat{\partial} \psi - \nonumber \\
 && - \frac{i}{2} \bar{\Omega}^\mu \hat{\partial} \Omega_\mu + i
(\bar{\Omega}\gamma) (\partial\Omega) - \frac{i}{2}
(\bar{\Omega}\gamma) \hat{\partial} (\gamma\Omega) + \frac{i}{2}
\bar{\chi} \hat{\partial} \chi
\end{eqnarray}
it is not hard to check that the most general supertransformations
leaving sum of massless Lagrangians invariant have the form
(round brackets denote symmetrization):
\begin{eqnarray}
\delta \Phi_{\mu\nu} &=& - \frac{i}{\sqrt{2}} \sigma^{\alpha\beta}
\partial_\alpha ( \cos(\theta_2) h_{\beta(\mu} \gamma_{\nu)} -
\sin(\theta_2) \gamma_5 f_{\beta(\mu} \gamma_{\nu)} ) \eta \nonumber
\\
\delta h_{\mu\nu} &=& \sqrt{2} \cos(\theta_2) (\bar{\Phi}_{\mu\nu}
\eta) + i \sin(\theta_2) (\bar{\Psi}_{(\mu} \gamma_{\nu)} \eta) \\
\delta f_{\mu\nu} &=& \sqrt{2} \sin(\theta_2) (\bar{\Phi}_{\mu\nu}
\gamma_5 \eta) + i \cos(\theta_2) (\bar{\Psi}_{(\mu} \gamma_{\nu)}
\gamma_5 \eta) \nonumber \\
\delta \Psi_\mu &=& - \sigma^{\alpha\beta} \partial_\alpha (
\sin(\theta_2) h_{\beta\mu} + \cos(\theta_2) f_{\beta\mu} \gamma_5 )
\eta \nonumber
\end{eqnarray}
for the supermultiplets containing spin 2 fields and
\begin{eqnarray}
\delta \Omega_\mu &=& - \frac{i}{2\sqrt{2}} \sigma^{\alpha\beta}
(\cos(\theta_1) A_{\alpha\beta} - \sin(\theta_1) \gamma_5
B_{\alpha\beta}) \gamma_\mu \eta \nonumber \\
\delta A_\mu &=& \sqrt{2} \cos(\theta_1) (\bar{\Omega}_\mu \eta) + i
\sin(\theta_1) (\bar{\psi} \gamma_\mu \eta) \\
\delta B_\mu &=& \sqrt{2} \sin(\theta_1) (\bar{\Omega}_\mu \gamma_5
\eta) + i \cos(\theta_1) (\bar{\psi} \gamma_\mu \gamma_5 \eta)
\nonumber \\
\delta \psi &=& - \frac{1}{2} \sigma^{\alpha\beta} ( \sin(\theta_1)
A_{\alpha\beta} + \cos(\theta_1) \gamma_5 B_{\alpha\beta}) \eta
\nonumber
\end{eqnarray}
for those with (axial)vector ones. The last supermultiplets is simple:
$$
\delta \chi = - i \gamma^\mu \partial_\mu (\varphi + \gamma_5 \pi)
\eta, \qquad \delta \varphi = (\bar{\chi} \eta), \qquad \delta \pi =
(\bar{\chi} \gamma_5 \eta)
$$

To construct massive supermultiplet we have to add mass terms for all
fields as well as appropriate corrections to fermionic
supertransformations. In this, the most important question is which
lower spin fields play the roles of Goldstone ones and have to be
eaten to make main gauge fields massive. For the bosonic fields
(taking into account parity conservation) the choice is unambiguous:
vector $A_\mu$ and scalar $\varphi$ fields for tensor field
$h_{\mu\nu}$ and axial-vector $B_\mu$ and pseudo-scalar $\pi$ --- for
pseudo-tensor $f_{\mu\nu}$. Thus bosonic mass terms will be:
\begin{eqnarray}
\frac{1}{m} {\cal L}_1 &=& \sqrt{2} [h^{\mu\nu} \partial_\mu A_\nu - h
(\partial A)] - \sqrt{3} A^\mu \partial_\mu \varphi + 
\sqrt{2} [f^{\mu\nu} \partial_\mu B_\nu - f (\partial B)] -
\sqrt{3} B^\mu \partial_\mu \pi \nonumber \\
\frac{1}{m^2} {\cal L}_2 &=& - \frac{1}{2} (h^{\mu\nu} h_{\mu\nu} -
 h^2) - \sqrt{\frac{3}{2}} h \varphi + \varphi^2 - 
\frac{1}{2} (f^{\mu\nu} h_{\mu\nu} -  f^2) - \sqrt{\frac{3}{2}}
f \pi + \pi^2
\end{eqnarray}
But for fermions we have two spin 3/2 and two spin 1/2 fields and
there is no evident choice. Thus we introduce the most general mass
terms for the fermions:
\begin{eqnarray}
\frac{1}{m} {\cal L}_m &=& - \frac{1}{2} \bar{\Phi}^{\mu\nu}
\Phi_{\mu\nu} + (\bar{\Phi}\gamma)^\mu (\gamma\Phi)_\mu + \frac{1}{4}
\bar{\Phi} \Phi - \nonumber \\
 && - i \alpha_1 [ \bar{\Phi}^{\mu\nu} \gamma_\mu \Psi_\nu 
 - \frac{1}{2} \bar{\Phi} (\gamma\Psi) ] - i \alpha_2 [
\bar{\Phi}^{\mu\nu} \gamma_\mu \Omega_\nu  - \frac{1}{2}
\bar{\Phi} (\gamma\Omega) ] + \nonumber \\
 && + a_1 [ \bar{\Psi}^\mu \Psi_\mu -  (\bar{\Psi}\gamma)
(\gamma\Psi) ] + a_2 [ \bar{\Omega}^\mu \Omega_\mu - 
(\bar{\Omega}\gamma) (\gamma\Omega) ] + a_3 [ \bar{\Psi}^\mu
\Omega_\mu -  (\bar{\Psi}\gamma) (\gamma\Omega)] + \nonumber \\
 && + i a_4 (\bar{\Psi}\gamma) \psi + i a_5 (\bar{\Psi}\gamma) \chi
+ i a_6 (\bar{\Omega}\gamma) \psi + i a_7 (\bar{\Omega}\gamma) \chi +
\nonumber \\
 && + a_8 \bar{\psi} \psi + a_9 \bar{\psi} \chi + a_{10} \bar{\chi}
\chi
\end{eqnarray}
and proceed with calculations. Cancellation of variations with one
derivative gives:
$$
\sin(\theta_2) = \cos(\theta_2) = \frac{1}{\sqrt{2}}
\quad \alpha_1 = \frac{1}{\sqrt{2}}
$$
$$
\sin(\theta_1) = \cos(\theta_1) = \frac{1}{\sqrt{2}}, \quad
\alpha_2 = \sqrt{2}
$$
$$
a_1 = - \frac{1}{4}, \quad a_3 = 1,
\quad a_4 = \sqrt{2}, \quad a_5 = 0
$$
$$
a_6 = a_2 \sqrt{2}, \quad a_7 = \sqrt{3},
\quad a_8 = 0, \quad a_9 = - \sqrt{6},
$$
while variations without derivatives give:
$$
a_2 = \frac{1}{2}, \qquad a_{10} = - \frac{1}{2}
$$
Resulting fermionic mass terms:
\begin{eqnarray}
\frac{1}{m} {\cal L}_m &=& - \frac{1}{2} \bar{\Phi}^{\mu\nu}
\Phi_{\mu\nu} + (\bar{\Phi}\gamma)^\mu (\gamma\Phi)_\mu + \frac{1}{4}
\bar{\Phi} \Phi - \nonumber \\
 && - \frac{i}{\sqrt{2}} \bar{\Phi}^{\mu\nu} \gamma_\mu \Psi_\nu 
 + \frac{i}{2\sqrt{2}} \bar{\Phi} (\gamma\Psi) - i \sqrt{2}
\bar{\Phi}^{\mu\nu} \gamma_\mu \Omega_\nu  + \frac{i}{\sqrt{2}}
\bar{\Phi} (\gamma\Omega) + \nonumber \\
 && - \frac{1}{4} \bar{\Psi}^\mu \Psi_\mu + \frac{1}{4}
(\bar{\Psi}\gamma) (\gamma\Psi) + \frac{1}{2} \bar{\Omega}^\mu
\Omega_\mu - \frac{1}{2} (\bar{\Omega}\gamma) (\gamma\Omega) + 
\bar{\Psi}^\mu \Omega_\mu - (\bar{\Psi}\gamma) (\gamma\Omega) +
\nonumber \\
 && + i \sqrt{2} (\bar{\Psi}\gamma) \psi + \frac{i}{\sqrt{2}}
(\bar{\Omega}\gamma) \psi + i \sqrt{3} (\bar{\Omega}\gamma) \chi -
\sqrt{6} \bar{\psi} \chi - \frac{1}{2} \bar{\chi} \chi
\end{eqnarray}
correspond to invariance of the Lagrangian (besides global
supertransformations) under three local spinor gauge transformations:
\begin{eqnarray}
\delta \Phi_{\mu\nu} &=& \partial_{(\mu} \xi_{\nu)} + \frac{im}{2}
\gamma_{(\mu} \xi_{\nu)} + \frac{m}{4\sqrt{2}} g_{\mu\nu} \xi_1 +
\frac{m}{2\sqrt{2}} g_{\mu\nu} \xi_2 \nonumber \\
\delta \Psi_\mu &=& \partial_\mu \xi_1 + \frac{m}{\sqrt{2}} \xi_\mu -
\frac{im}{4} \gamma_\mu \xi_1 + \frac{im}{2} \gamma_\mu \xi_2
\nonumber \\
\delta \Omega_\mu &=& \partial_\mu \xi_2 + m\sqrt{2} \xi_\mu +
\frac{im}{2} \gamma_\mu \xi_1 + \frac{im}{2} \gamma_\mu \xi_2   \\
\delta \psi &=& m\sqrt{2} \xi_1 + \frac{m}{\sqrt{2}} \xi_2
\qquad \delta \chi = m\sqrt{3} \xi_2 \nonumber
\end{eqnarray}
From these formula one can easily determine which combination of spin
3/2 fields $\Psi_\mu$ and $\Omega_\mu$ plays the role of Goldstone
field for spin 5/2 field $\Phi_{\mu\nu}$. Indeed, if one introduces
two orthogonal combinations:
$$
\tilde{\Psi}_\mu = \frac{1}{\sqrt{5}} \Psi_\mu + \frac{2}{\sqrt{5}}
\Omega_\mu, \quad \tilde{\Omega}_\mu = - \frac{2}{\sqrt{5}} \Psi_\mu
+ \frac{1}{\sqrt{5}} \Omega_\mu
$$
then after diagonalization of mass terms one finds that
$\tilde{\Psi}_\mu$ is a Goldstone field, while $\tilde{\Omega}_\mu$
--- physical field with the same mass as $\Phi_{\mu\nu}$.

Similar to the case of massive supermultiplet with superspin 1, both
mixing angles have been fixed: $\theta_1 = \theta_2 = \pi/4$ and
this, in turn, means that all bosonic fields enter through the complex
combinations only:
$$
H_{\mu\nu} = h_{\mu\nu} + \gamma_5 f_{\mu\nu}, \qquad
C_\mu = A_\mu + \gamma_5 B_\mu, \qquad z = \varphi + \gamma_5 \pi
$$
Introducing gauge covariant derivatives:
$$
\nabla_\mu H_{\alpha\beta} = \partial_\mu H_{\alpha\beta} -
\frac{m}{\sqrt{2}} C_\mu g_{\alpha\beta}, \qquad
\nabla_\mu z = \partial_\mu z - m \sqrt{3} C_\mu
$$
we can write final form of fermionic supertransformations as:
\begin{eqnarray}
\delta \Phi_{\mu\nu} &=& [ - \frac{i}{2} \sigma^{\alpha\beta}
\nabla_\alpha \bar{H}_{\beta(\mu} \gamma_{\nu)} - m H_{\mu\nu} +
\frac{m}{4} \gamma_{(\mu} (\gamma H)_{\nu)} + \frac{m}{4}
\sqrt{\frac{3}{2}} g_{\mu\nu} z ] \eta \nonumber \\ 
\delta \Psi_\mu &=& [ - \frac{1}{\sqrt{2}} \sigma^{\alpha\beta}
\nabla_\alpha H_{\beta\mu} - \frac{i m}{2\sqrt{2}} (\gamma H)_\mu 
 + \frac{1}{\sqrt{3}} \nabla_\mu z + \frac{im}{4\sqrt{3}} \gamma_\mu z
] \eta \nonumber \\
\delta \Omega_\mu &=& [ - \frac{i}{4} \sigma^{\alpha\beta}
\bar{C}_{\alpha\beta} \gamma_\mu + \frac{i m}{\sqrt{2}} (\gamma H)_\mu
 + \frac{1}{\sqrt{3}} \nabla_\mu z - \frac{im}{2\sqrt{3}} \gamma_\mu z
] \eta \\
\delta \psi &=& - \frac{1}{2\sqrt{2}} \sigma^{\alpha\beta}
C_{\alpha\beta} \eta \qquad \delta \chi = - i \gamma^\mu \nabla_\mu z
\eta \nonumber
\end{eqnarray}

Note also that due to complexification of bosonic fields the
Lagrangian and supertransformations are invariant under global axial
$U(1)_A$ symmetry, axial charges for all fields being:
\begin{center}
\begin{tabular}{|c|c|c|c|} \hline
field  & $\eta$ & $\Phi_{\mu\nu}$, $\Psi_\mu$, $\Omega_\mu$, $\psi$,
$\chi$ & $H_{\mu\nu}$, $C_\mu$, $z$ \\
\hline $q_A$ & +1 & 0 & --1 \\ \hline
\end{tabular}
\end{center}

\section{Superspin 5/2}

Our next example --- massive supermultiplet with superpin 5/2. It
also contains four massive fields: with spin 3, 5/2, 5/2' and 2 and
in the massless limit it should reduce to six massless
supermultiplets:
$$
\left( \begin{array}{ccc}  & 3 & \\ 5/2 &  & 5/2 \\  & 2 &
\end{array} \right) \quad \Rightarrow \quad
\left( \begin{array}{c} 3 \\ 5/2 \end{array} \right) \oplus
\left( \begin{array}{c} 5/2 \\ 2 \end{array} \right) \oplus
\left( \begin{array}{c} 2 \\ 3/2 \end{array} \right) \oplus
\left( \begin{array}{c} 3/2 \\ 1 \end{array} \right) \oplus
\left( \begin{array}{c} 1 \\ 1/2 \end{array} \right) \oplus
\left( \begin{array}{c} 1/2 \\ 0,0' \end{array} \right)
$$
By analogy with all previous cases we will take into account possible
mixing for bosonic tensor and vector fields, so we will start with
the following structure of massless supermultiplets:
$$
\left( \begin{array}{c} \Phi_{\mu\nu\lambda} \\ \Psi_{\mu\nu}
\end{array} \right) \quad \oplus \quad
\left( \begin{array}{ccc}  & \Omega_{\mu\nu} & \\ h_{\mu\nu} & &
f_{\mu\nu} \\  & \Psi_\mu & \end{array} \right) \quad \oplus \quad
\left( \begin{array}{ccc}  & \Omega_\mu & \\ A_\mu &  & B_\mu \\  &
\psi & \end{array} \right) \quad \oplus \quad
\left( \begin{array}{c} \chi \\ z \end{array} \right)
$$
So we introduce sum of the massless Lagrangians for bosonic fields:
\begin{eqnarray}
{\cal L}_0 &=& - \frac{1}{2} \partial^\rho \Phi^{\mu\nu\lambda}
\partial_\rho \Phi_{\mu\nu\lambda} + \frac{3}{2} (\partial
\Phi)^{\mu\nu} (\partial \Phi)_{\mu\nu} - 3 (\partial \Phi)^{\mu\nu}
\partial_\mu \Phi_\nu + \frac{3}{2} \partial^\mu \Phi^\nu \partial_\mu
\Phi_\nu + \frac{3}{4} (\partial \Phi)^2 \nonumber \\
 && + \frac{1}{2} \partial^\alpha h^{\mu\nu} \partial_\alpha
h_{\mu\nu} - (\partial h)^\mu (\partial h)_\mu + (\partial h)^\mu
\partial_\mu h - \frac{1}{2} \partial^\mu h \partial_\mu h
- \frac{1}{4} A_{\mu\nu}{}^2  + \frac{1}{2} (\partial_\mu
\varphi)^2 \nonumber \\
 && + \frac{1}{2} \partial^\alpha f^{\mu\nu} \partial_\alpha
f_{\mu\nu} - (\partial f)^\mu (\partial f)_\mu + (\partial f)^\mu
\partial_\mu f - \frac{1}{2} \partial^\mu f \partial_\mu f
- \frac{1}{4} B_{\mu\nu}{}^2  + \frac{1}{2} (\partial_\mu \pi)^2 
\end{eqnarray}
as well as sum of the massless Lagrangians for fermionic fields:
\begin{eqnarray}
{\cal L}_0 &=& \frac{i}{2} \bar{\Psi}^{\mu\nu} \hat{\partial}
\Psi_{\mu\nu} - 2 i (\bar{\Psi}\gamma)^\mu (\partial\Psi)_\mu 
+ i (\bar{\Psi}\gamma)^\mu \hat{\partial} (\gamma\Psi)_\mu +
i (\bar{\Psi}\gamma\partial) \Psi - \frac{i}{4}
\bar{\Psi} \hat{\partial} \Psi + \nonumber \\
 && + \frac{i}{2} \bar{\Omega}^{\mu\nu} \hat{\partial}
\Omega_{\mu\nu} - 2 i (\bar{\Omega}\gamma)^\mu (\partial\Omega)_\mu 
+ i (\bar{\Omega}\gamma)^\mu \hat{\partial} (\gamma\Omega)_\mu +
i (\bar{\Omega}\gamma\partial) \Omega - \frac{i}{4}
\bar{\Omega} \hat{\partial} \Omega - \nonumber \\
 && - \frac{i}{2} \bar{\Psi}^\mu \hat{\partial} \Psi_\mu + i
(\bar{\Psi}\gamma) (\partial\Psi) - \frac{i}{2}
(\bar{\Psi}\gamma) \hat{\partial} (\gamma\Psi) + \frac{i}{2}
\bar{\psi} \hat{\partial} \psi - \nonumber \\
 && - \frac{i}{2} \bar{\Omega}^\mu \hat{\partial} \Omega_\mu + i
(\bar{\Omega}\gamma) (\partial\Omega) - \frac{i}{2}
(\bar{\Omega}\gamma) \hat{\partial} (\gamma\Omega) + \frac{i}{2}
\bar{\chi} \hat{\partial} \chi
\end{eqnarray}
and start with the following global supertransformations:
\begin{equation}
\delta \Phi_{\mu\nu\lambda} = i ( \bar{\Psi}_{(\mu\nu}
\gamma_{\lambda)} \eta) \qquad 
\delta \Psi_{\mu\nu} =  [ - \sigma^{\alpha\beta}
\partial_\alpha \Phi_{\beta\mu\nu} + \frac{1}{4} \partial_{(\mu}
\gamma_{\nu)} (\gamma \Phi) ] \eta
\end{equation}
for the supermultiplet $(3,5/2)$,
\begin{eqnarray}
\delta \Omega_{\mu\nu} &=& - \frac{i}{\sqrt{2}} \sigma^{\alpha\beta}
\partial_\alpha ( \cos(\theta_2) h_{\beta(\mu} \gamma_{\nu)} -
\sin(\theta_2) \gamma_5 f_{\beta(\mu} \gamma_{\nu)} ) \eta \nonumber
\\
\delta h_{\mu\nu} &=& \sqrt{2} \cos(\theta_2) (\bar{\Omega}_{\mu\nu}
\eta) + i \sin(\theta_2) (\bar{\Psi}_{(\mu} \gamma_{\nu)} \eta) \\
\delta f_{\mu\nu} &=& \sqrt{2} \sin(\theta_2) (\bar{\Omega}_{\mu\nu}
\gamma_5 \eta) + i \cos(\theta_2) (\bar{\Psi}_{(\mu} \gamma_{\nu)}
\gamma_5 \eta) \nonumber \\
\delta \Psi_\mu &=& - \sigma^{\alpha\beta} \partial_\alpha (
\sin(\theta_2) h_{\beta\mu} + \cos(\theta_2) f_{\beta\mu} \gamma_5 )
\eta \nonumber
\end{eqnarray}
for the mixed $(5/2,2)$ and $(2,3/2)$ supermultiplets,
\begin{eqnarray}
\delta \Omega_\mu &=& - \frac{i}{2\sqrt{2}} \sigma^{\alpha\beta}
(\cos(\theta_1) A_{\alpha\beta} - \sin(\theta_1) \gamma_5
B_{\alpha\beta}) \gamma_\mu \eta \nonumber \\
\delta A_\mu &=& \sqrt{2} \cos(\theta_1) (\bar{\Omega}_\mu \eta) + i
\sin(\theta_1) (\bar{\psi} \gamma_\mu \eta) \\
\delta B_\mu &=& \sqrt{2} \sin(\theta_1) (\bar{\Omega}_\mu \gamma_5
\eta) + i \cos(\theta_1) (\bar{\psi} \gamma_\mu \gamma_5 \eta) 
\nonumber \\
\delta \psi &=& - \frac{1}{2} \sigma^{\alpha\beta} ( \sin(\theta_1)
A_{\alpha\beta} + \cos(\theta_1) \gamma_5 B_{\alpha\beta}) \eta
\nonumber
\end{eqnarray}
for the mixed $(3/2,1)$ and $(1,1/2)$ supermultiplets and
$$
\delta \chi = - i \gamma^\mu \partial_\mu (\varphi + \gamma_5 \pi)
\eta, \qquad \delta \varphi = (\bar{\chi} \eta), \qquad \delta \pi =
(\bar{\chi} \gamma_5 \eta)
$$
for the last one.

By analogy with the superspin 3/2 case, we will assume that fermionic
mass terms are Dirac ones:
\begin{eqnarray}
\frac{1}{m} {\cal L}_m &=& - \bar{\Psi}^{\mu\nu} \Omega_{\mu\nu} +
2 (\bar{\Psi}\gamma)^\mu (\gamma\Omega)_\mu + \frac{1}{2}
\bar{\Psi} \Omega + \nonumber \\
 && + i \sqrt{\frac{5}{2}} [ - \bar{\Psi}^{\mu\nu} \gamma_\mu
\Psi_\nu  + \frac{1}{2} \bar{\Psi} (\gamma\Psi) -
\bar{\Omega}^{\mu\nu} \gamma_\mu \Omega_\nu  + \frac{1}{2}
\bar{\Omega} (\gamma\Omega) ] + \nonumber \\
 && + \frac{3}{2} \bar{\Psi}^\mu \Omega_\mu - \frac{3}{2} (\bar{\Psi}
\gamma) (\gamma \Omega) + 2 i (\bar{\Psi} \gamma) \psi + 2 i
(\bar{\Omega} \gamma) \chi - 3 \bar{\psi} \chi
\end{eqnarray}
where all coefficients are completely fixed by the requirement that
the Lagrangian has to be invariant not only under the global
supertransformations, but under four (by the number of fermionic
gauge fields) spinor gauge transformations:
\begin{eqnarray*}
\delta \Psi_{\mu\nu} &=& \partial_{(\mu} \xi_{\nu)} + \frac{im}{2}
\gamma_{(\mu} \eta_{\nu)} + \frac{m}{4} \sqrt{\frac{5}{2}} g_{\mu\nu}
\xi_1, \quad \delta \Omega_{\mu\nu} = \partial_{(\mu} \eta_{\nu)} + 
\frac{im}{2} \gamma_{(\mu} \xi_{\nu)} + \frac{m}{4} \sqrt{\frac{5}{2}}
g_{\mu\nu} \xi_2,  \\
\delta \Psi_\mu &=& \partial_\mu \xi_1 + m\sqrt{\frac{5}{2}} \xi_\mu +
im \frac{3}{4} \gamma_\mu \xi_2, \quad \delta \Omega_\mu =
\partial_\mu \xi_2 + m\sqrt{\frac{5}{2}} \eta_\mu
+ im \frac{3}{4} \gamma_\mu \xi_1, \\
\delta \psi &=& 2m \xi_1, \qquad \delta \chi = 2m \xi_2 
\end{eqnarray*}

As for the bosonic fields, here the roles of the fields are evident
(again taking into account parity conservation): we need tensor
$h_{\mu\nu}$, vector $A_\mu$ and scalar $\varphi$ fields to make spin
3 field $\Phi_{\mu\nu\lambda}$ massive, while pseudo-tensor
$f_{\mu\nu}$ field needs to eat axial-vector $B_\mu$ and pseudo-scalar
$\pi$ fields. So the bosonic mass terms are also completely fixed:
\begin{eqnarray}
\frac{1}{m} {\cal L}_1 &=& \sqrt{3} [ - \Phi^{\mu\nu\lambda}
\partial_\mu h_{\nu\lambda} + 2 \Phi^\mu (\partial h)_\mu -
\frac{1}{2} \Phi^\mu \partial_\mu h ] + \sqrt{5} [ h^{\mu\nu}
\partial_\mu A_\nu - h (\partial A) ] - \sqrt{6} A^\mu \partial_\mu
\varphi + \nonumber \\
 && + \sqrt{2} [f^{\mu\nu} \partial_\mu B_\nu - f (\partial B)] -
\sqrt{3} B^\mu \partial_\mu \pi \\
\frac{1}{m^2} {\cal L}_2 &=& \frac{1}{2} \Phi^{\mu\nu\lambda}
\Phi_{\mu\nu\lambda} - \frac{3}{2} \Phi^\mu \Phi_\mu + \frac{3}{4} h^2
+ \frac{\sqrt{15}}{2} \Phi^\mu A_\mu - \frac{3}{4} A_\mu{}^2 - 
\sqrt{\frac{15}{2}} h \varphi + \frac{5}{2} \varphi^2 - \nonumber \\
 && - \frac{1}{2} (f^{\mu\nu} f_{\mu\nu} -  f^2) - \sqrt{\frac{3}{2}}
f \pi + \pi^2
\end{eqnarray}

Now we require that the whole Lagrangian be invariant under global
supertransformations with appropriate corrections to fermionic
transformations. This fixes both mixing angles:
$$
\sin(\theta_2) = \sqrt{\frac{5}{6}}, \quad
\cos(\theta_2) = \frac{1}{\sqrt{6}}, \quad
\sin(\theta_1) = \sqrt{\frac{2}{3}}, \quad
\cos(\theta_1) = \frac{1}{\sqrt{3}}
$$
and gives the following form of additional terms for fermionic
supertransformations:
\begin{eqnarray}
\frac{1}{m} \delta \Psi_{\mu\nu} &=& [ - \frac{\sqrt{3}}{4}
\gamma_{(\mu} (\gamma h)_{\nu)} - \sqrt{\frac{5}{3}} f_{\mu\nu}
\gamma_5 + \frac{1}{4} \sqrt{\frac{5}{3}} \gamma_{(\mu}
(\gamma f)_{\nu)} \gamma_5 ] \eta \nonumber \\
\frac{1}{m} \delta \Omega_{\mu\nu} &=& i [ (\gamma \Phi)_{\mu\nu}
+ \frac{1}{8} g_{\mu\nu} (\gamma \Phi) - \nonumber \\
 && - \frac{1}{4} \sqrt{\frac{5}{3}} \gamma_{(\mu} A_{\nu)} +
\frac{1}{4} \sqrt{\frac{5}{3}} g_{\mu\nu} \hat{A} - \frac{1}{2}
\sqrt{\frac{5}{6}} \gamma_{(\mu} B_{\nu)} \gamma_5 + \frac{1}{2}
\sqrt{\frac{5}{6}} g_{\mu\nu} \hat{B} \gamma_5 ] \eta \nonumber \\
\frac{1}{m} \delta \Psi_\mu &=& [ \frac{1}{8} \sqrt{\frac{5}{2}}
\gamma_\mu (\gamma \Phi) - \frac{1}{2\sqrt{6}} A_\mu -
\frac{5}{2\sqrt{6}} \gamma_\mu \hat{A} - \frac{5}{2\sqrt{3}} B_\mu
\gamma_5 - \frac{1}{2\sqrt{3}} \gamma_\mu \hat{B} \gamma_5 ] \eta \\
\frac{1}{m} \delta \Omega_\mu &=& i [ \frac{3}{2} \sqrt{\frac{5
}{6}}(\gamma h)_\mu + \frac{1}{2} \sqrt{\frac{3}{2}} (\gamma f)_\mu
\gamma_5 - \gamma_\mu \varphi - \gamma_\mu \gamma_5 \pi ] \eta 
\nonumber \\
\frac{1}{m} \delta \psi &=& [ - \varphi - 2 \gamma_5 \pi ] \eta \qquad
\frac{1}{m} \delta \chi = i [ \sqrt{6} \hat{A} + \sqrt{3} \hat{B}
\gamma_5 ] \eta \nonumber
\end{eqnarray}
The complete supertransformations for fermionic fields could be
simplified by introduction of gauge invariant derivatives;
$$
\nabla_\mu h_{\alpha\beta} = \partial_\mu h_{\alpha\beta} -
\frac{m\sqrt{5}}{2} A_\mu g_{\alpha\beta}, \qquad \nabla_\mu
\varphi = \partial_\mu \varphi - m \sqrt{6} A_\mu
$$
$$
\nabla_\mu f_{\alpha\beta} = \partial_\mu f_{\alpha\beta} -
\frac{m}{\sqrt{2}} B_\mu g_{\alpha\beta}, \qquad \nabla_\mu \pi =
\partial_\mu \pi - m \sqrt{3} B_\mu
$$
This time bosonic fields do not combine into complex combinations,
but due to the fact that fermionic mass terms are Dirac ones the
Lagrangian and supertransformations are invariant under global axial
$U(1)_A$ transformations, provided axial charges of all fields are
assigned as follows:
\begin{center}
\begin{tabular}{|c|c|c|c|} \hline
field  & $\eta$, $\Psi_{\mu\nu}$, $\Psi_\mu$, $\psi$ &
$h_{\mu\nu}$, $f_{\mu\nu}$, $A_\mu$, $B_\mu$, $\varphi$, $\pi$ &
$\Omega_{\mu\nu}$, $\Omega_\mu$, $\chi$  \\
\hline $q_A$ & +1 & 0 & --1 \\ \hline
\end{tabular}
\end{center}

\section{Massive particles}

All our previous and subsequent calculations heavily depend on
the gauge invariant description of massive high spin particles. For
reader convenience and to make paper self-contained we will give here
gauge invariant formulations for massive particles with arbitrary
integer \cite{KZ97,Zin01} and half-integer \cite{Met06} spins. We
restrict ourselves to flat $d=4$ Minkowski space but all results could
be easily generalized to the case of (A)dS space with arbitrary
dimension $d$.

\subsection{Integer spin}

The simplest way to describe massless bosonic field with arbitrary
spin $s$ is to use completely symmetric rank $s$ tensor
$\Phi_{(\alpha_1 \alpha_2\dots\alpha_s)}$ which is double traceless.
In what follows we will use condensed notations where index denotes
just number of free indices and not the indices themselves. For
example, the tensor field itself will be denoted as $\Phi_s$, it's
contraction with derivative as $(\partial \Phi)_{s-1}$, it's trace as
$\tilde{\Phi}_{s-2}$ and so on. As we will see this does not lead to
any ambiguities then working with free Lagrangians quadratic in
fields. In these notations the Lagrangian for massless particles of
arbitrary spin $s$ could be written as:
\begin{eqnarray}
{\cal L}_0 &=& (-1)^s [ \frac{1}{2} \partial^\mu \Phi^s \partial_\mu
\Phi^s - \frac{s}{2} (\partial \Phi)^{s-1} (\partial \Phi)^{s-1}
- \frac{s(s-1)}{4} \partial^\mu \tilde{\Phi}^{s-2} \partial_\mu
\tilde{\Phi}^{s-2} + \nonumber \\
 && + \frac{s(s-1)}{2} (\partial \Phi)^{s-1} \partial_{(1}
\tilde{\Phi}_{s-2)} - \frac{s(s-1)(s-2)}{8} (\partial
\tilde{\Phi})^{s-3} (\partial \tilde{\Phi})^{s-3}  ]
\end{eqnarray}
where $\tilde{\tilde{\Phi}} = 0$. This Lagrangian is invariant under
the following gauge transformations:
$$
\delta_0 \Phi_s = \partial_{(1} \xi_{s-1)}, \qquad
\tilde{\xi}_{s-3} = 0
$$
where parameter $\xi_{s-1}$ is completely symmetric traceless tensor
of rank $s-1$.

To construct gauge invariant Lagrangian for massive particle which
has correct (i.e. with right number of physical degrees of freedom)
massless limit, we start with the sum of massless Lagrangians with
$0 \le k \le s$:
\begin{eqnarray}
{\cal L}_0 &=& \sum_{k=0}^s (-1)^k [ \frac{1}{2} \partial^\mu \Phi^k
\partial_\mu \Phi^k - \frac{k}{2} (\partial \Phi)^{k-1} (\partial
\Phi)^{k-1} - \frac{k(k-1)}{4} \partial^\mu \tilde{\Phi}^{k-2}
\partial_\mu \tilde{\Phi}^{k-2} \nonumber \\
 && + \frac{k(k-1)}{2} (\partial \Phi)^{k-1} \partial_{(1}
\tilde{\Phi}_{k-2)} - \frac{k(k-1)(k-2)}{8} (\partial
\tilde{\Phi})^{k-3} (\partial \tilde{\Phi})^{k-3}  ]
\end{eqnarray}
Then we add the following cross terms with one derivative as well as
mass terms without derivatives:
\begin{eqnarray}
\frac{1}{m} {\cal L}_1 &=& \sum_{k=1}^s (-1)^k a_k [ (\partial
\Phi)^{k-1} \Phi_{k-1} + (k-1) \tilde{\Phi}^{k-2} (\partial
\Phi)_{k-2} + \frac{(k-1)(k-2)}{4} (\partial \tilde{\Phi})^{k-3}
\tilde{\Phi}_{k-3} ] \nonumber \\
\frac{1}{m^2} {\cal L}_2 &=& \sum_{k=0}^s (-1)^k [ d_k \Phi^k \Phi_k +
e_k \tilde{\Phi}^{k-2} \tilde{\Phi}_{k-2} + f_k \tilde{\Phi}^{k-2}
\Phi_{k-2} ]
\end{eqnarray}
and try to achieve gauge invariance with the help of appropriate
corrections to gauge transformations:
$$
\frac{1}{m} \delta \Phi_k = \alpha_k \xi_k + \beta_k g_{(2}
\xi_{k-2)}
$$
Straightforward but lengthy calculations give a number of algebraic
equations on the unknown coefficients which could be solved (and this
is non-trivial because we obtain overdetermined system of equations)
and give us:
$$
\alpha_k{}^2 = \frac{(s-k)(s+k+1)}{2(k+1)^2}, \quad
\beta_{k+1} = \frac{k+1}{2k} \alpha_k, \quad
0 \le k \le s-1
$$
$$
a_k = - \sqrt{\frac{(s-k+1)(s+k)}{2}}, \quad 
d_k = \frac{(s-k-1)(s+k+2)}{4(k+1)}
$$
$$
e_k = \frac{k(k-1)}{16(k+1)} [ (s-k+2)(s+k-1) + 6]
$$
$$
f_k = - \frac{1}{4} \sqrt{(s-k+2)(s+k-1)(s-k+1)(s+k)}
$$

\subsection{Half-integer spin}

For the description of massless spin $s+1/2$ particles we will use
completely symmetric rank $s$ tensor-spinor $\Psi_s$ such that 
$(\gamma \tilde{\Psi})_{s-3} = 0$ (in the same condensed notations as
before). Then Lagrangain for such field could be written as:
\begin{eqnarray}
{\cal L}_0 &=& i (-1)^s [ \frac{1}{2} \bar{\Psi}^s \hat{\partial}
\Psi^s - s (\bar{\Psi} \gamma)^{s-1} (\partial \Psi)^{s-1} +
\frac{s}{2} (\bar{\Psi} \gamma)^{s-1} \hat{\partial} (\gamma
\Psi)^{s-1} +  \nonumber \\
 && \qquad + \ \frac{s(s-1)}{2} (\bar{\Psi} \gamma \partial)^{s-2}
\tilde{\Psi}^{s-2} - \frac{s(s-1)}{8} \tilde{\bar{\Psi}}^{s-2}
\hat{\partial} \tilde{\Psi}^{s-2} ]
\end{eqnarray}
and is invariant under the following gauge transformations:
$$
\delta_0 \Psi_s = \partial_{(1} \xi_{s-1)}, \qquad (\gamma \xi) = 0,
$$
where gauge parameter $\xi_{s-1}$ is a $\gamma$-traceless
tensor-spinor of rank $s-1$.

Once again we start with the sum of massless Lagrangians with $0 \le
k \le s$:
\begin{eqnarray}
{\cal L}_0 &=& \sum_{k=0}^{s} i (-1)^k [ \frac{1}{2} \bar{\Psi}^k
\hat{\partial} \Psi^k - k (\bar{\Psi} \gamma)^{k-1} (\partial
\Psi)^{k-1} + \frac{k}{2} (\bar{\Psi} \gamma)^{k-1} \hat{\partial}
(\gamma \Psi)^{k-1} + \nonumber \\
 && \qquad + \ \frac{k(k-1)}{2} (\bar{\Psi} \gamma \partial)^{k-2}
\tilde{\Psi}^{k-2} - \frac{k(k-1)}{8} \tilde{\bar{\Psi}}^{k-2}
\hat{\partial} \tilde{\Psi}^{k-2} ]
\end{eqnarray}

To combine all these massless fields into one massive particle we
have to add the following mass terms:
\begin{eqnarray}
{\cal L}_m &=& \sum_{k=0}^{s} (-1)^k \left\{ - \frac{s+1}{2(k+1)}
[ \bar{\Psi}^k \Psi^k - k (\bar{\Psi}\gamma)^{k-1} (\gamma\Psi)^{k-1}
- \frac{k(k-1)}{4} \tilde{\bar{\Psi}}^{k-2} \tilde{\Psi}^{k-2} ] +
\right. \nonumber \\
 && \qquad \qquad \left. - i c_k [ (\bar{\Psi}\gamma)^{k-1} \Psi^{k-1}
- \frac{k-1}{2} \tilde{\bar{\Psi}}^{k-2} (\gamma\Psi)^{k-2} ] \right\}
\end{eqnarray}
and corresponding corrections to gauge transformations:
$$
\delta \Psi_k = \alpha_k \xi_k + i \beta_k \gamma_{(1} \xi_{k-1)} +
\rho_k g_{(2} \xi_{k-2)}
$$
Then total Lagrangian will be gauge invariant provided:
$$
c_k = \sqrt{\frac{(s+1)^2 - k^2}{2}}, \qquad
\alpha_k = \frac{c_{k+1}}{k+1}, \qquad
\beta_k = \frac{s+1}{2k(k+1)}, \qquad
\rho_k = \frac{c_k{}}{2k}
$$

\section{Massless supermultiplets}

It is not easy to find in the recent literature the explicit component
form of massless supermultiplets with arbitrary superspin
\cite{Cur79}, so for completeness we will give their short description
here. As we have already seen on the lower superspin cases,
supermultiplets with integer and half-integer superspins have
different structure and have to be considered separately.

{\bf (s, s+1/2)}.  Supermultiplet with integer superspin $s$
contains bosonic spin s field and fermionic spin $s+1/2$ one. In this
and in two subsequent sections we will use the same condensed
notations as in the previous one. By analogy with superspins 1 and 2
supermultiplets, we start with the following ansatz for the
supertransformations:
$$
\delta \Psi_s = i \alpha_1 \sigma^{\mu\nu} \partial_\mu
\Phi_{\nu(s-1} \gamma_{1)} \eta \qquad
\delta \Phi_s = \beta ( \bar{\Psi}_s \eta )
$$
Indeed, calculating variations of the sum of two massless Lagrangians
one can see that most of variations cancel, provided one set 
$\alpha_1 = - \frac{\beta}{2}$. The residue:
\begin{eqnarray*}
\delta {\cal L} &=& - (-1)^s \beta \frac{(s-1)(s-2)}{4} [ 2
(\bar{\Psi} \partial \partial)^{s-2} \tilde{\Phi}_{s-2} -
\tilde{\bar{\Psi}}^{s-2} \partial^2 \tilde{\Phi}_{s-2} - (s-2)
(\tilde{\bar{\Psi}} \partial)^{s-3} (\partial \tilde{\Phi})_{s-3} - \\
 && - 2 (\bar{\Psi} \gamma \partial)^{s-2} \hat{\partial}
\tilde{\Phi}_{s-2} + 2 (\bar{\Psi} \gamma \partial \partial)^{s-3}
(\gamma \tilde{\Phi})_{s-3} - (\tilde{\bar{\Psi}}
\partial)^{s-3} \hat{\partial} (\gamma \tilde{\Phi})_{s-3} ]
\end{eqnarray*}
contains terms with $\tilde{\Phi}_{s-2}$ only, so we proceed by
adding to fermionic supertransformations one more term:
$$
\delta' \Psi_s = i \alpha_2 \partial_{(1} \gamma_1
\tilde{\Phi}_{s-2)}
$$
Then the choice $\alpha_2 = \frac{(s-1)(s-2)\beta}{4s}$ leaves us
with:
$$
\delta {\cal L} = - (-1)^s \beta \frac{(s-1)(s-2)}{4} [ 2 (\bar{\Psi}
\gamma \partial \partial)^{s-3} (\gamma \tilde{\Phi})_{s-3} -
(\tilde{\bar{\Psi}} \partial)^{s-3} \hat{\partial} (\gamma
\tilde{\Phi})_{s-3} ]
$$
where the only terms are whose with $(\gamma \tilde{\Phi})_{s-3}$. So
we make one more (last) correction to supertransformations:
$$
\delta'' \Psi_s = i \alpha_3 g_{(2} \partial_1 (\gamma
\tilde{\Phi})_{s-3)}
$$
and obtain full invariance with $\alpha_3 = -
\frac{(s-1)(s-2)\beta}{4s}$. To fix concrete normalization we
will use closure of the superalgebra. Calculating the commutator
of two supertransformations we obtain:
$$
[ \delta_1, \delta_2 ] \Phi_s = - i \beta^2 (\bar{\eta}_2 \gamma^\nu
\eta_1) \partial_\mu \Phi_s + \dots
$$
where dots mean ``up to gauge transformation''. So we set
$\beta = \sqrt{2}$ and our final result looks like:
\begin{eqnarray}
\delta \Psi_s &=& - \frac{i}{\sqrt{2}} \sigma^{\mu\nu} \partial_\mu
\bar{\Phi}_{\nu(s-1} \gamma_{1)} \eta + \frac{i(s-1)(s-2)}{2\sqrt{2}s}
[ \partial_{(1} \gamma_1 \tilde{\Phi}_{s-2)} - g_{(2} \partial_1
(\gamma \tilde{\Phi})_{s-3)} ] \eta \nonumber \\
\delta \Phi_s &=& \sqrt{2} ( \bar{\Psi}_s \eta) 
\end{eqnarray}

{\bf (s+1/2, s+1)}. Half-integer superspin multiplet contains
fermionic spin $s+1/2$ fields and bosonic spin $s+1$ one. Again by
analogy with lower superspin case we will make the following ansatz
for supertransformations:
$$
\delta \Psi_s = \alpha_1 \sigma^{\mu\nu} \partial_\mu \Phi_{\nu(s)}
\eta, \qquad \delta \Phi_{s+1} = i \beta (\bar{\Psi}_{(s} \gamma_{1)}
\eta)
$$
This time most of the variations cancel if one set $\alpha_1 = -
\beta$ leaving us with:
\begin{eqnarray*}
\delta {\cal L} &=& i (-1)^s \beta \frac{s(s-1)}{4} [ 2 (\bar{\Psi}
\partial \partial)^{s-2} (\gamma \tilde{\Phi})_{s-2} -
\tilde{\bar{\Psi}}^{s-2} \partial^2 (\gamma \tilde{\Phi})_{s-2} - \\
 && \qquad \qquad - 2 (\bar{\Psi} \gamma \partial)^{s-2}
\hat{\partial} (\gamma \tilde{\Phi})_{s-2} - (s-2) (\tilde{\bar{\Psi}}
\partial)^{s-3} (\gamma \partial \tilde{\Phi})_{s-3} ]
\end{eqnarray*}
Then the full invariance could be achieved with the following
correction to supertransformations:
$$
\delta' \Psi_s = \alpha_2 \partial_{(1} \gamma_1
(\gamma \tilde{\Phi})_{s-2)}
$$
provided $\alpha_2 = \frac{s-1}{4}$.
To check the closure of superalgebra and to choose normalization we
calculate commutator of two supertransformations:
$$
[ \delta_1, \delta_2 ] \Phi_{s+1} = - 2 i \beta^2 (\bar{\eta}_2
\gamma^\mu \eta_1) \partial_\mu \Phi_{s+1} + \dots
$$
Then our choice will be $\beta = 1$ and our final form:
$$
\delta \Psi_s = - \sigma^{\mu\nu} \partial_\mu \Phi_{\nu(s)} \eta +
\frac{s-1}{4} \partial_{(1} \gamma_1 (\gamma \tilde{\Phi})_{s-2)}
\eta,
\qquad \delta \Phi_{s+1} = i ( \bar{\Psi}_{(s} \gamma_{1)} \eta )
$$
Note that starting with superspin 2 the structure of
supertransformations are defined up to possible field dependent gauge
transformations and our choice differs from that of \cite{Cur79}. It
makes no difference for massless theories but for massive case the
structure of corrections for fermionic supertransformations depends
on the choice made.

\section{Integer superspin}

Now, having in our disposal gauge invariant description of massive
particles with arbitrary (half-)integer spins, known form of
supertransformations for massless arbitrary superspin supermultiplets
and concrete examples of massive supermultiplets with lower
superspins, we are ready to construct massive arbitrary
superspin supermultiplets. As we have seen, integer and half-integer
cases have different structures and have to be considered separately.

In this section we consider massive supermultiplet with integer
superspin. Such supermultiplet also contains four massive fields: two
bosonic spin s fields (with opposite parity) and fermionic spin
(s+1/2) and (s-1/2) ones. Calculating total number of physical
degrees of freedom and taking into account possible mixing of
supermultiplets containing bosonic fields with equal spins and
opposite parity, we start with the following structure of massless
supermultiplets:
$$
\left( \begin{array}{ccc}  & \Phi_s & \\ A_s & & B_s \\ & \Psi_{s-1} &
\end{array} \right) \qquad \Rightarrow \qquad \sum_{k=1}^s \quad
\left( \begin{array}{ccc}  & \Phi_k & \\ A_k & & B_k \\ & \Psi_{k-1} &
\end{array} \right) \quad \oplus \quad 
\left( \begin{array}{c} \Phi_0 \\ z \end{array} \right)
$$
By analogy with superspin 1 and 2 cases, we will assume that all
bosonic fields enter through the complex combinations
$C_k = A_k + i B_k$ only (so that all possible mixing angles are
fixed and equal $\pi/4$). Thus we choose the following form of
supertransformations for massless supermultiplets with $1 \le k \le
s$:
\begin{eqnarray}
\delta \Phi_k &=& - \frac{i}{2} \sigma^{\mu\nu} \partial_\mu
\bar{C}_{\nu(k-1} \gamma_{1)} \eta + \frac{i(k-1)(k-2)}{4k} [
\partial_{(1} \gamma_1 \tilde{C}_{k-2)} - g_{(2} \partial_1 (\gamma
\tilde{C})_{k-3)} ] \eta \nonumber \\
\delta \bar{C}_k &=& 2 ( \bar{\Phi}_k \eta) + i\sqrt{2}
(\bar{\Psi}_{(k-1} \gamma_{1)} \eta) \\
\delta \Psi_{k-1} &=& - \frac{1}{\sqrt{2}} \sigma^{\mu\nu}
\partial_\mu C_{\nu(k-1)} \eta + \frac{k-2}{4\sqrt{2}} \partial_{(1}
\gamma_1 (\gamma \tilde{C})_{k-3)} \eta \nonumber
\end{eqnarray}
and also
$$
\delta \Phi_0 = - i \hat{\partial} z \eta, \qquad
\delta \bar{z} = 2 ( \bar{\Phi}_0 \eta)
$$
As a result of our assumption mass terms for bosonic fields are
completely fixed:
\begin{eqnarray}
{\cal L}_1 &=& \sum_{k=0}^s (-1)^k c_k [ \bar{C}^k \partial_{(1}
C_{k-1)} - (k-1) \tilde{\bar{C}}^{k-2} (\partial C)_{k-2} +
\frac{(k-1)(k-2)}{4} (\tilde{\bar{C}}^{k-2} \partial_{(1}
\tilde{C}_{k-3)} + h.c.)] \nonumber \\
{\cal L}_2 &=& \sum_{k=0}^s (-1)^k [ d_k \bar{C}^k C_k + e_k
\tilde{\bar{C}}^{k-2} \tilde{C}_{k-2} + f_k ( \tilde{\bar{C}}^{k-2}
C_{k-2} + h.c. ) ]
\end{eqnarray}
where
$$
c_k = \frac{1}{2} \sqrt{\frac{(s+k)(s-k+1)}{2}}, \qquad
d_k = \frac{(s-k-1)(s+k+2)}{4(k+1)}
$$
$$
e_k = \frac{k(k-1)}{16(k+1)} [ (s-k+2)(s+k-1) + 6]
$$
$$
f_k = - \frac{1}{8} \sqrt{(s-k+2)(s+k-1)(s-k+1)(s+k)}
$$
As for the fermionic mass terms, apriori we don't have any
restrictions on them so we have to consider the most general possible
form:
\begin{eqnarray}
\frac{1}{m} {\cal L}_m &=& \sum_{k=0}^s (-1)^k \left[ 
 a_{1k} [ \bar{\Phi}^k \Phi_k - k (\bar{\Phi} \gamma)^{k-1} (\gamma
\Phi)_{k-1} - \frac{k(k-1)}{4} \tilde{\bar{\Phi}}^{k-2}
\tilde{\Phi}_{k-2}] + \right. \nonumber \\
 && + a_{2k} [ \bar{\Phi}^k \Psi_k - k (\bar{\Phi} \gamma)^{k-1}
(\gamma \Psi)_{k-1} - \frac{k(k-1)}{4} \tilde{\bar{\Phi}}^{k-2}
\tilde{\Psi}_{k-2}] + \nonumber \\
 && + a_{3k} [ \bar{\Psi}^k \Psi_k - k (\bar{\Psi} \gamma)^{k-1}
(\gamma \Psi)_{k-1} - \frac{k(k-1)}{4} \tilde{\bar{\Psi}}^{k-2}
\tilde{\Psi}_{k-2}] + \nonumber \\
 && + i b_{1k} [ (\bar{\Phi} \gamma)^{k-1} \Phi_{k-1} -
\frac{k-1}{2} \tilde{\bar{\Phi}}^{k-2} (\gamma \Phi)_{k-2} ] + 
\nonumber \\
 && + i b_{2k} [ (\bar{\Phi} \gamma)^{k-1} \Psi_{k-1} -
\frac{k-1}{2} \tilde{\bar{\Phi}}^{k-2} (\gamma \Psi)_{k-2} ] +
\nonumber \\
 && + i b_{3k} [ (\bar{\Psi} \gamma)^{k-1} \Phi_{k-1} -
\frac{k-1}{2} \tilde{\bar{\Psi}}^{k-2} (\gamma \Phi)_{k-2} ] +
\nonumber \\
 && + \left. i b_{4k} [ (\bar{\Psi} \gamma)^{k-1} \Psi_{k-1} -
\frac{k-1}{2} \tilde{\bar{\Psi}}^{k-2} (\gamma \Psi)_{k-2} ] \right]
\end{eqnarray}
Where:
$$
a_{1s} = - \frac{1}{2}, \qquad a_{2s} = a_{3s} = b_{3s} = b_{4s} = 0
$$
The requirement that total Lagrangian be invariant under
(appropriately corrected) supertransformations gives:
$$
a_{1k} = - \frac{1}{2}, \qquad
a_{2k} = - \frac{2\sqrt{2}}{k+1} c_{k+1}, \qquad
a_{3k} = \frac{k}{2(k+1)}
$$
$$
b_{1k} = - 2 c_k, \quad b_{2k} = - \frac{1}{\sqrt{2}}, \quad
b_{3k} = 0, \quad b_{4k} = - 2 c_{k+1}
$$
In this, additional terms for fermionic supertransformations look
like:
\begin{eqnarray}
\frac{1}{m} \delta' \Phi_k &=& \frac{2ic_{k+1}}{k+1} [ (\gamma C)_k +
\frac{k(k-1)}{4(k+1)} \gamma_{(1} \tilde{C}_{k-1)} -
\frac{(k-1)^2(2k+1)}{8k(k+1)} g_{(2} (\gamma \tilde{C})_{k-2)} ] -
\nonumber \\
 && - C_k + \frac{k-1}{2k} \gamma_{(1} (\gamma C)_{k-1} -
\frac{(k-1)(k-2)}{8k^2} g_{(2} \tilde{C}_{k-2)} - \nonumber \\
 && - \frac{ic_k}{k} [ \gamma_{(1} C_{k-1)} - g_{(2} (\gamma
C)_{k-2)} ]  \\
\frac{1}{m} \delta' \Psi_k &=& \frac{k c_{k+2}}{2\sqrt{2}(k+1)}
\gamma_{(1} (\gamma \tilde{C})_{k-1)} - \nonumber \\
 && - \frac{ik}{\sqrt{2}(k+1)}  [ (\gamma C)_k - \frac{k(k-1)}{4(k+1)}
\gamma_{(1} \tilde{C}_{k-1)} + \frac{(k-1)(3k+1)}{8k(k+1)} g_{(2}
(\gamma \tilde{C})_{k-2)} ] - \nonumber \\
 && - \frac{\sqrt{2} c_{k+1}}{k+1} [ C_k + \frac{1}{k} \gamma_{(1}
(\gamma C)_{k-1)} + \frac{(k-1)(k-2)}{4k^2} g_{(2} \tilde{C}_{k-2)} ]
\end{eqnarray}
Here the supertransformations for $\Phi_k$ field contain terms with
$C_{k+1}$, $C_k$ and $C_{k-1}$ fields in the first, second and third
lines correspondingly, while that of $\Psi_k$ contain terms with
$C_{k+2}$, $C_{k+1}$ and $C_k$ fields.

\section{Half-integer superspin}

Next we turn to the half-integer superspin case. This time we have
two fermionic spin (s+1/2) fields and bosonic ones with spins (s+1)
and s. Usual reasoning on physical degrees of freedom and possible
mixings leads us to the following structure of massless
supermultiplets we will start with:
$$
\left( \begin{array}{ccc}  & A_{s+1} & \\ \Phi_s & & \Psi_s \\ &
B_s & \end{array} \right) \quad \Rightarrow \quad
\left( \begin{array}{c} A_{s+1} \\ \Psi_s \end{array} \right) \quad
\oplus \quad \sum_{k=1}^s \quad
\left( \begin{array}{ccc}  & \Phi_k & \\ A_k & & B_k \\ & \Psi_{k-1} &
\end{array} \right) \quad \oplus \quad 
\left( \begin{array}{c} \Phi_0 \\ z \end{array} \right)
$$
We see that this structure is rather similar to that of integer
superspin case. The main difference (besides the presence of
$A_{s+1}$, $\Psi_s$ supermultiplet) comes from the mixing of bosonic
fields. We have no reasons to suggest that all mixing angles could be
fixed from the very beginning so we have to consider the most general
possibility here. Let us denote:
$$
C_k = \cos(\theta_k) A_k + \gamma_5 \sin(\theta_k) B_k, \qquad
D_k = \sin(\theta_k) A_k + \gamma_5 \cos(\theta_k) B_k
$$
In these notations supertransformations for massless supermultiplets
could be written as follows. Highest supermultiplet:
$$
\delta A_{s+1} = i ( \bar{\Psi}_{(s} \gamma_{1)} \eta ) \qquad
\delta \Psi_s = - \sigma^{\mu\nu} \partial_\mu A_{\nu(s)} \eta +
\frac{s-1}{4} \partial_{(1} \gamma_1 (\gamma \tilde{A})_{s-2)} \eta
$$
Main set ($1 \le k \le s$):
\begin{eqnarray}
\delta \Phi_k &=& - \frac{i}{\sqrt{2}} \sigma^{\mu\nu} \partial_\mu
\bar{C}_{\nu(k-1} \gamma_{1)} \eta + \frac{i(k-1)(k-2)}{2\sqrt{2}k} [
\partial_{(1} \gamma_1 \tilde{C}_{k-2)} - g_{(2} \partial_1 (\gamma
\tilde{C})_{k-3)} ] \eta \nonumber \\
\delta A_k &=& \sqrt{2} \cos(\theta_k) ( \bar{\Phi}_k \eta) + i
\sin(\theta_k) ( \bar{\Psi}_{(k-1} \gamma_{1)} \eta) \\
\delta B_k &=& \sqrt{2} \sin(\theta_k) ( \bar{\Phi}_k \gamma_5 \eta) +
i \cos(\theta_k) ( \bar{\Psi}_{(k-1} \gamma_{1)} \gamma_5 \eta) 
\nonumber \\
\delta \Psi_{k-1} &=& - \sigma^{\mu\nu} \partial_\mu D_{\nu(k-1)}
\eta + \frac{k-2}{4} \partial_{(1} \gamma_1 (\gamma \tilde{D})_{k-3}
\eta \nonumber
\end{eqnarray}
and the last supermultiplet:
$$
\delta \Phi_0 = - i \hat{\partial} z \eta, \qquad
\delta \bar{z} = 2 ( \bar{\Phi}_0 \eta)
$$
By analogy with superspins 3/2 and 5/2 cases we will assume that
fermionic mass terms are Dirac ones. This immediately gives:
\begin{eqnarray}
{\cal L}_f &=& \sum_{k=0}^s (-1)^k \left\{ - \frac{s+1}{k+1} [
\bar{\Psi}^k \Phi_k - k (\bar{\Psi} \gamma)^{k-1} (\gamma \Phi)_{k-1}
- \frac{k(k-1)}{4} \tilde{\bar{\Psi}}^{k-2} \tilde{\Phi}_{k-2} ] -
\right. \nonumber \\
 && - \left. i  c_k [ (\bar{\Psi} \gamma)^{k-1} \Psi_{k-1} -
\frac{k-1}{2} \tilde{\bar{\Psi}}^{k-2} (\gamma \Psi)_{k-2} +
( \Psi \rightarrow \Phi ) ] \right\}
\end{eqnarray}
where:
$$
c_k = \sqrt{\frac{(s+k+1)(s-k+1)}{2}}
$$
The choice for the bosonic mass terms (taking into account parity)
is also unambiguous:
\begin{eqnarray}
{\cal L}_1 &=& \sum_{k=0}^{s+1} (-1)^k a_k [ A^k \partial_{(1}
A_{k-1)} - (k-1) \tilde{A}^{k-2} (\partial A)_{k-2} +
\frac{(k-1)(k-2)}{4} \tilde{A}^{k-2} \partial_{(1} \tilde{A}_{k-3)} ]
+ \nonumber \\
 && \sum_{k=0}^{s} (-1)^k b_k [ B^k \partial_{(1}
B_{k-1)} - (k-1) \tilde{B}^{k-2} (\partial B)_{k-2} +
\frac{(k-1)(k-2)}{4} \tilde{B}^{k-2} \partial_{(1} \tilde{B}_{k-3)} ]
\end{eqnarray}
for the terms with one derivative, where: 
$$
a_k = \sqrt{\frac{(s+k+1)(s-k+2)}{2}}, \qquad b_k =
\sqrt{\frac{(s+k)(s-k+1)}{2}}
$$
and the following terms without derivatives:
\begin{eqnarray}
\frac{1}{m^2} {\cal L}_2 &=& \sum_{k=0}^{s+1} (-1)^k [ \hat{d}_k
A^k A_k + \hat{e}_k \tilde{A}^{k-2} \tilde{A}_{k-2} + \hat{f}_k
\tilde{A}^{k-2} A_{k-2} ] + \nonumber \\
 && + \sum_{k=0}^s (-1)^k [ d_k B^k B_k + e_k \tilde{B}^{k-2}
\tilde{B}_{k-2} + f_k \tilde{B}^{k-2} B_{k-2} ]
\end{eqnarray}
Here:
$$
\hat{d}_k = \frac{(s-k)(s+k+3)}{4(k+1)}, \quad
\hat{e}_k = \frac{k(k-1)}{16(k+1)} [ (s-k+3)(s+k) + 6]
$$
$$
\hat{f}_k = - \frac{1}{4} \sqrt{(s-k+3)(s+k)(s-k+2)(s+k+1)}
$$
$$
d_k = \frac{(s-k-1)(s+k+2)}{4(k+1)}, \quad
e_k = \frac{k(k-1)}{16(k+1)} [ (s-k+2)(s+k-1) + 6]
$$
$$
f_k = - \frac{1}{4} \sqrt{(s-k+2)(s+k-1)(s-k+1)(s+k)}
$$
Note that hatted coefficients differ from the unhatted ones by
replacement $s \rightarrow s+1$.

Now we require that total Lagrangian be invariant under the
supertransformations. First of all this fixes all mixing angles:
$$
\sin(\theta_k) = \sqrt{\frac{s+k+1}{2(s+1)}}, \quad
\cos(\theta_k) = \sqrt{\frac{s-k+1}{2(s+1)}}
$$
and gives us additional terms for fermionic supertransformations:
\begin{eqnarray*}
\frac{1}{m} \delta' \Psi_k &=& \alpha_1 A_k + \alpha_2
\gamma_{(1}(\gamma A)_{k-1)} + \alpha_3 g_{(2} \tilde{A}_{k-2)} + \\
 && + \beta_1 B_k + \beta_2 \gamma_{(1}(\gamma B)_{k-1)} + \beta_3
g_{(2} \tilde{B}_{k-2)} + \\
 && + \frac{k c_{k+1}}{4(k+1)} [
\sin(\theta_{k+2}) \gamma_{(1} (\gamma \tilde{A})_{k-1)} +
\cos(\theta_{k+1}) \gamma_{(1} (\gamma \tilde{B})_{k-1)} ] \\
\frac{1}{m} \delta' \Phi_k &=& \alpha_4 (\gamma A)_k + \alpha_5
\gamma_{(1} \tilde{A}_{k-1)} + \alpha_6 g_{(2} (\gamma
\tilde{A})_{k-2)} + \\
 && + \beta_4 (\gamma B)_k + \beta_5 \gamma_{(1} \tilde{B}_{k-1)} +
\beta_6 g_{(2} (\gamma \tilde{B})_{k-2)} - \\
 && - \frac{a_k}{k\sqrt{2}} \cos(\theta_k) [ \gamma_{(1} A_{k-1)} -
g_{(2} (\gamma A)_{k-2} ] - \\
 && - \frac{b_k}{k\sqrt{2}} \sin(\theta_k) [\gamma_{(1} B_{k-1)} -
g_{(2} (\gamma B)_{k-2)} ]
\end{eqnarray*}
Where:
$$
\alpha_1 = - \frac{s-k}{\sqrt{2}(k+1)} \cos(\theta_k), \quad
\alpha_2 = - \frac{k^2+s+k+1}{\sqrt{2}k(k+1)} \cos(\theta_k)
$$
$$
\alpha_3 = - \frac{(s+1)(k-1)(k-2)}{4\sqrt{2}k^2(k+1)} \cos(\theta_k)
$$
$$
\beta_1 = - \frac{s+k+2}{\sqrt{2}(k+1)} \sin(\theta_k), \quad
\beta_2 = \frac{k^2-s+k-1}{\sqrt{2}k(k+1)} \sin(\theta_k)
$$
$$
\beta_3 = - \frac{(s+1)(k-1)(k-2)}{4\sqrt{2}k^2(k+1)} \sin(\theta_k)
$$

$$
\alpha_4 = \frac{s+1}{k+1} \sin(\theta_{k+1}), \qquad
\alpha_5 = \frac{k(k-1)(s-k)}{4(k+1)^2} \sin(\theta_{k+1})
$$
$$
\alpha_6 = \frac{(k-1) [(k+1)(s+1) - 2k^2 (s-k)]}{8k(k+1)^2}
\sin(\theta_{k+1})
$$
$$
\beta_4 = \frac{s+1}{k+1} \cos(\theta_{k+1}), \qquad
\beta_5 = \frac{k(k-1)(s+k+2)}{4(k+1)^2} \cos(\theta_{k+1})
$$
$$
\beta_6 = \frac{(k-1) [(k+1)(s+1) - 2k^2 (s+k+2)]}{8k(k+1)^2}
\cos(\theta_{k+1})
$$

We have explicitely checked that (rather complicated) formulas from
this and previous sections correctly reproduce all lower superspins
results.

\section*{Conclusion}

Thus, using supersymmetric generalization of gauge invariant
description for massive particles, we managed to show that all massive
$N=1$ supermultiplets could be constructed out of appropriate set of
massless ones. In this, in spite of large number of fields involved,
all calculations are pretty straightforward and mainly
combinatorical. Certainly, using gauge invariance one can fix the
gauge where all but four physical massive fields are equal to zero.
But in this case all supertransformations must be supplemented with
field dependent gauge transformations restoring the gauge. So the
structure of resulting supertransformation become very complicated
and will contain higher derivative terms.


\begin{thebibliography}{10}

\bibitem{Zin83}
Yu.~M. Zinoviev
{\it "Gauge invariant description of massive high spin particles"}
Preprint 83-91, IHEP, Protvino, 1983.

\bibitem{KZ97}
S.~M. Klishevich, Yu.~M. Zinoviev
{\it "On electromagnetic interaction of massive spin-2 particle",}
Phys. Atom. Nucl. {\bf 61} (1998) 1527, arXiv:hep-th/9708150.

\bibitem{Zin01}
Yu.~M. Zinoviev
{\it "On Massive High Spin Particles in (A)dS",} arXiv:hep-th/0108192.

\bibitem{AGS02}
N.~Arkani-Hamed, H.~Georgi, M.~D. Schwartz
{\it "Effective Field Theory for Massive Gravitons and Gravity in
Theory Space",}
Ann. Phys. {\bf 305} (2003) 96, arXiv:hep-th/0210184.

\bibitem{Ham05}
S.~Hamamoto
{\it "Possible Nonlinear Completion of Massive Gravity",}
Prog. Theor. Phys. {\bf 114} (2006) 1261, arXiv:hep-th/0505194.

\bibitem{BHR05}
M.~Bianchi, P.~J. Heslop, F.~Riccioni
{\it "More on La Grande Bouffe",}
JHEP {\bf 08} (2005) 088, arXiv:hep-th/0504156.

\bibitem{HW05}
K.~Hallowell, A.~Waldron
{\it "Constant Curvature Algebras and Higher Spin Action Generating
  Functions",}
Nucl. Phys. {\bf B724} (2005) 453, arXiv:hep-th/0505255.

\bibitem{BK05}
I.~L. Buchbinder, V.~A. Krykhtin
{\it "Gauge invariant Lagrangian construction for massive bosonic
higher spin fields in D dimensions",}
Nucl. Phys. {\bf B727} (2005) 537, arXiv:hep-th/0505092.

\bibitem{BKL06}
I.~L. Buchbinder, V.~A. Krykhtin, P.~M. Lavrov
{\it "Gauge invariant Lagrangian formulation of higher spin massive
bosonic field theory in AdS space",}
Nucl. Phys. {\bf B762} (2007) 344, arXiv:hep-th/0608005.

\bibitem{Met06}
R.~R. Metsaev
{\it "Gauge invariant formulation of massive totally symmetric
fermionic fields in (A)dS space",}
Phys. Lett. {\bf B643} (2006) 205-212, arXiv:hep-th/0609029.

\bibitem{Zin06}
Yu.~M. Zinoviev
{\it "On massive spin 2 interactions",}
Nucl. Phys. {\bf B770} (2007) 83, arXiv:hep-th/0609170.

\bibitem{Met06a}
R.~R. Metsaev
{\it "Gravitational and higher-derivative interactions of massive spin
5/2 field in (A)dS space",} arXiv:hep-th/0612279.

\bibitem{BKR07}
I.~L. Buchbinder, V.~A. Krykhtin, A.~A. Reshetnyak
{\it "BRST approach to Lagrangian construction for fermionic higher
spin fields in (A)dS space",} arXiv:hep-th/0703049.

\bibitem{BGPL02}
I.L. Buchbinder, Jr. S.~J.~Gates, J.~Phillips, W.~D. Linch
{\it "New 4D, N = 1 Superfield Theory: Model of Free Massive
Superspin-3/2 Multiplet",}
Phys. Lett. {\bf B535} (2002) 280-288, arXiv:hep-th/0201096.

\bibitem{Zin02}
Yu.~M. Zinoviev
{\it "Massive Spin-2 Supermultiplets",} arXiv:hep-th/0206209.

\bibitem{BGLP02}
I.~L. Buchbinder, S.~J.~Gates Jr, W.D.~Linch III, J.~Phillips
{\it "Dynamical Superfield Theory of Free Massive Superspin-1
Multiplet",}
Phys. Lett. {\bf B549} (2002) 229-236, arXiv:hep-th/0207243.

\bibitem{GSS04}
T.~Gregoire, M.~D. Schwartz, Y.~Shadmi
{\it "Massive Supergravity and Deconstruction",}
JHEP {\bf 0407} (2004) 029, arXiv:hep-th/0403224.

\bibitem{GK05}
S.~J. Gates and  S.~M. Kuzenko
{\it "4D, N =1 Higher Spin Gauge Superfields and Quantized Twistors",}
JHEP {\bf 0510} (2005) 008, arXiv:hep-th/0506255.

\bibitem{BGKP05}
I.~L. Buchbinder, S.~J.~Gates Jr., S.~M. Kuzenko, J.~Phillips
{\it "Massive 4D, N = 1 Superspin 1 \& 3/2 Multiplets and Dualities",}
JHEP {\bf 0502} (2005) 056, arXiv:hep-th/0501199.

\bibitem{GKTM06}
S.~J.~Gates Jr., S.~M. Kuzenko, G.~Tartaglino-Mazzucchelli
{\it "New massive supergravity multiplets",} arXiv:hep-th/0610333.

\bibitem{Zin07}
Yu.~M. Zinoviev
{\it "Massive supermultiplets with spin 3/2",}
arXiv:hep-th/0703118.

\bibitem{Cur79}
T.~Curtright
{\it "Massless field supermultiplets with arbitrary spin",}
Phys. Lett. {\bf B85} (1979) 219.

\end{thebibliography}
\end{document}